\documentclass[aps,prl,twocolumn,groupedaddress,article,superscriptaddress,showpacs,amsmath,amssymb]{revtex4-1}

\usepackage{graphicx}
\usepackage{dcolumn}
\usepackage{bm}

\newcommand{\Ppost}{P_{\mathrm{post}}}

\newcommand{\Psrs}{P_{\mathrm{SRS}}}

\newcommand{\Pthr}{P_{\mathrm{thr}}}
\newcommand{\Rmax}{R_{\mathrm{max}}}

\newcommand{\degree}{$^{\circ}\,$}

\newcommand{\be}{\begin{equation}}
\newcommand{\ee}{\end{equation}}
\newcommand{\bea}{\begin{eqnarray}}
\newcommand{\eea}{\end{eqnarray}}
\newcommand{\bi}{\begin{itemize}}
\newcommand{\ei}{\end{itemize}}
\newcommand{\ben}{\begin{enumerate}}
\newcommand{\een}{\end{enumerate}}

\newcommand{\bw}{\begin{widetext}}
\newcommand{\ew}{\end{widetext}}
\newcommand{\bwe}{\begin{widetext}\begin{eqnarray}}
\newcommand{\ewe}{\end{eqnarray}\end{widetext}}

\begin{document}

\title{Raman backscatter as a remote laser power sensor in high-energy-density plasmas}

\author{J. D. Moody}
\email{moody4@llnl.gov}
\author{D. J. Strozzi}
\author{L. Divol}
\author{P. Michel}
\author{H. F. Robey}
\author{S. LePape}
\author{J. Ralph}
\author{J. S. Ross}
\affiliation{Lawrence Livermore National Laboratory, Livermore, California 94551, USA}
\author{S. H. Glenzer}
\affiliation{LCLS, Stanford, California, USA}
\author{R. K. Kirkwood}
\author{O. L. Landen}
\author{B. J. MacGowan}
\affiliation{Lawrence Livermore National Laboratory, Livermore, California 94551, USA}
\author{A. Nikroo}
\affiliation{General Atomics, San Diego, California, USA}
\author{E. A. Williams}
\affiliation{Lawrence Livermore National Laboratory, Livermore, California 94551, USA}

\date{\today}

\begin{abstract}
Stimulated Raman backscatter (SRS) is used as a remote sensor to quantify the instantaneous laser power after transfer from outer to inner cones that cross in a National Ignition Facility (NIF) gas-filled hohlraum plasma.  By matching SRS between a shot reducing outer vs a shot reducing inner power we infer that $\sim$ half of the incident outer-cone power is transferred to inner cones, for the specific time and wavelength configuration studied.  This is the first instantaneous non-disruptive measure of power transfer in an indirect drive NIF experiment using optical measurements.
\end{abstract}


\maketitle

Advances in experimental science often are enabled by new diagnostic capabilities.  These may include increased accuracy, higher dynamic range, and accessibility to previously unavailable data records.  Intense laser-plasma interaction (LPI) studies have a well-documented history of such diagnostic-assisted advances \cite{froula,hutchinson,glenzer,kritcher,labaune}.  Additionally, improvements in diagnostics for characterizing the laser \cite{haynam} have led to greater experimental reproducibility and better comparisons with theoretical models.

An ongoing challenge to improving LPI understanding is measurement accessibility.  For example, laser and x-ray probe access in cylindrical hohlraums used for indirect drive inertial fusion \cite{lindl} is often limited to regions just outside of the Laser Entrance Holes (LEHs).  Experimenters have sometimes cut probe access holes into the hohlraum \cite{glenzer_hohl}, but this may alter the local plasma conditions and affect the measurement in an unknown way.  It would be beneficial to develop a probe for LPI that could be used in places inaccessible by standard laser probes.  Recent examples of this use parametrically scattered light to determine plasma and LPI properties \cite{froula_two,rosen}.

In this article we demonstrate a novel use of stimulated Raman backscatter (SRS) as a remote probe for laser power measurement in a NIF hohlraum region not accessible by standard probes.  We use this remote probe to detect the instantaneous power transferred between crossing laser beams \cite{moody_nature,michel_xb2010,femto_power_trans,michel_xb,michel_sat}.  Power transfer occurs through a 3-wave mixing process (akin to stimulated Brillouin scattering - SBS) in the low density LEH plasma where the laser beams overlap.  Time-averaged power transfer has been inferred in previous experiments from x-ray emission measurements of imploded capsule core symmetry \cite{meezan} or from laser hot-spots on the hohlraum wall \cite{michel_3color}.  More direct time-resolved transfer measurements are important for a quantitative understanding of the time-dependent cross-beam transfer physics as well as the ICF implosion dynamics.  This technique may be applicable in other systems with limited probe laser accessibility.

The experiments are performed with the NIF laser \cite{moses2005} using an indirect drive ignition hohlraum target \cite{haan,robey_one,robey_two}.  This target is a gold cylinder ($\sim$1 cm long and $\sim$ 0.5 cm diameter), cryogenically cooled to 21.5 K and filled with He gas at a density of 0.96 mg/cc.  The 192 NIF laser beams are positioned around a spherical target chamber, in four separate cones on each hemisphere.  The two inner cones are at 23.5\degree and 30\degree and the two outers cones are at 44.5\degree and 50\degree relative to the hohlraum axis.  The laser beams enter the target chamber in 2x2 groups of 4 beams called quads.  The experiments are done with about 1.1 MJ of laser energy and a peak laser power of 350 TW.  The outer-cone quads have an intensity at best focus of $8.1 \times 10^{\rm 14}$ W/cm$^2$ and the inners of $3.4 \times 10^{\rm 14}$ W/cm$^2$.

The most important point of this Letter stems from the data in Fig.~2.  Here we show two laser experiments which, at $t = 22$ ns, have the same plasma conditions, the same measured SRS (1 TW), a measured pump laser power in one case (7 TW) but an unknown pump power in the other (4 TW $+$ power-transfer).  Since the two incident laser powers give the same SRS in the same plasma, the two incident powers must be the same.  Thus, 7 TW $=$ 4 TW $+$ power-transfer.  This provides the key for using this technique to determine the instantaneous cross-beam power transfer.  Note that this is the power transfer at one specific point in time ($t = 22$ ns $\pm 0.1$ ns).  We require no additional knowledge, e.g.\ the intensity dependence of the SRS or the time-dependence of the plasma conditions.  The SRS spectrum in Fig.~3 establishes that the relevant plasma conditions in the two experiments are the same at $t = 22$ ns.  Figure 4 presents additional {\it secondary} data from several tries to get the two shots shown in Fig.~2.  These data provide an empirical SRS power scaling in the NIF hohlraum plasma (at one time), depicted in Fig.~5.

\begin{figure}[htbp]
   \centering
  \includegraphics[width=3.2in]{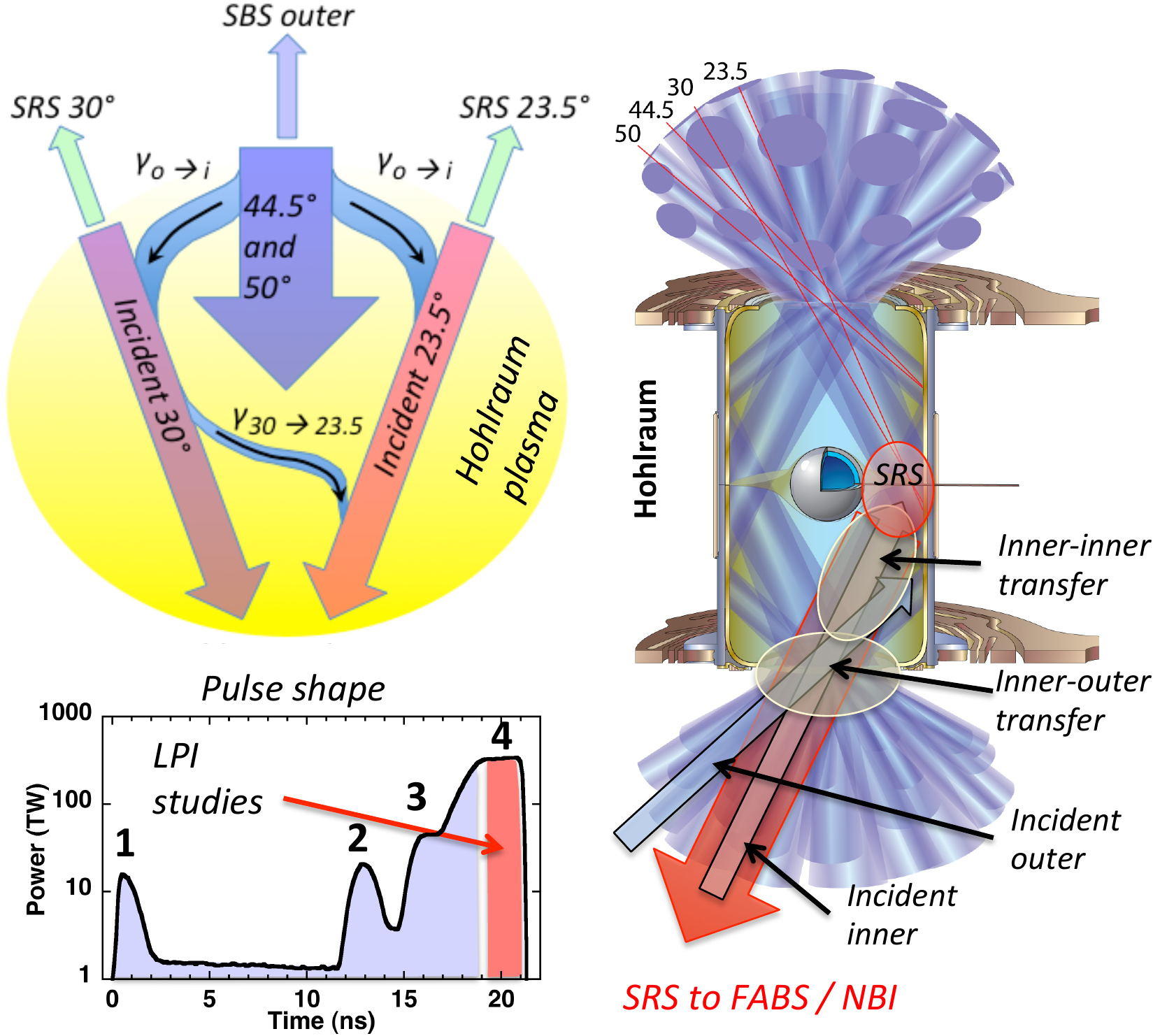} 
   \caption{Sketch of the hohlraum target used for these measurements.  Also shown is a typical laser pulse shape - the high-intensity part of the pulse is labeled with the number "4."  Power-transfer between the different beam groups for the "three-color" NIF configuration is shown in a sketch to the left the hohlraum.}
   \label{fig:fig_one}
\end{figure}

Figure \ref{fig:fig_one} shows the key aspects of the experiment.  The direction of power transfer for the "three-color" laser configuration (described below) is shown in the upper-left.  A sketch of the hohlraum showing the different beam cones and the direction of the measured SRS backscatter is on the right.  The pulse shape is shown in the lower-left.  SRS backscattered power is measured on a 30\degree quad using a time-resolved Near Backscatter Imager (NBI) \cite{moodyrsi}; the time-resolved SRS spectrum is measured with a Full-Aperture Backscatter System (FABS).  The FABS samples the portion of the SRS backscattered into the beam optics and may not show the same time-history as the total SRS measurement.   Measurement error for SRS power is $\pm$ 20\%.  The plasma density and temperature vary throughout the hohlraum but are approximately $n_e/n_{\rm cr} = 0.1$ ($n_{\rm cr} = 9 \times 10^{21}$ cm$^{-3}$) and $T_e = 2.5$ keV in the region where inner-cone SRS originates.  SRS gain estimates show that peak gains occur for Langmuir wavenumber $k \lambda_{\rm Debye} \approx 0.3$ \cite{rosen,trivel}.

The NIF laser is configured as "three-color" in these experiments \cite{michel_3color}.  The 44.5\degree and 50\degree (outer-cone) quads have the shortest wavelength $\lambda_{out}=$1052.43 nm (before frequency tripling), the 30\degree inner cone has an intermediate wavelength $\lambda_{30}=\lambda_{out}+7.3$ \AA, and the 23.5\degree inner cone has the longest wavelength $\lambda_{23}=\lambda_{30}+1.2$ \AA.  The sketch to the right of the target in Fig.~ \ref{fig:fig_one} shows that power transfers from the shorter to longer wavelength cones.  Power that transfers from the outer cones to both inner cones affects the imploded core's polar shape, while power transfer from the 30\degree to the 23.5\degree cones affects its azimuthal shape.  The inner-beam backscatter is observed to increase as the time-averaged power transfer to the inners increases, indicating backscatter occurs after power transfer, i.e.\ deeper in the hohlraum.

The solution to the 1D coupled mode equations for power transfer from quad (a) to quad (b), in the limit of small gain, is
\begin{equation}
\begin{aligned}
P_a^{(1)} & = P_a^{(0)} \times [1 - \gamma_{a \rightarrow b} P_b^{(0)}], \\
P_b^{(1)} & = P_b^{(0)} \times [1 + \gamma_{a \rightarrow b} P_a^{(0)}].
\end{aligned}
\label{eq:one}
\end{equation}
The superscripts refer to the power before (0) and after (1) transfer. $\gamma_{a \rightarrow b}$ describes the power transfer rate which depends on wavelength tuning, plasma conditions, and beam properties like overlap geometry.  We express the total, cone-to-cone coupling in terms of single quad powers (recall there are twice as many quads on the NIF outer vs. inner cones).  Transfer rates important for the three-color NIF configuration are: $\gamma_{o \rightarrow i}$ and $\gamma_{30 \rightarrow 23.5}$ and are shown in Fig.~\ref{fig:fig_one}.  Transfer from the outer to inner cones occurs in the LEH where these cones overlap.  We assume the same transfer coefficient $\gamma_{o \rightarrow i}$ applies to both the 23.5\degree and 30\degree cones for the large wavelength separation considered; this is consistent with transfer calculations and symmetry measurements.  Transfer between the two inner cones occurs over a longer overlap region extending between the LEH and the SRS scattering location.  Gain estimates in plasmas simulated for these experiments show that SRS occurs mostly where the inner beams pass between the capsule and hohlraum wall. 

\begin{figure}[htbp]
   \centering
  \includegraphics[width=3.2in]{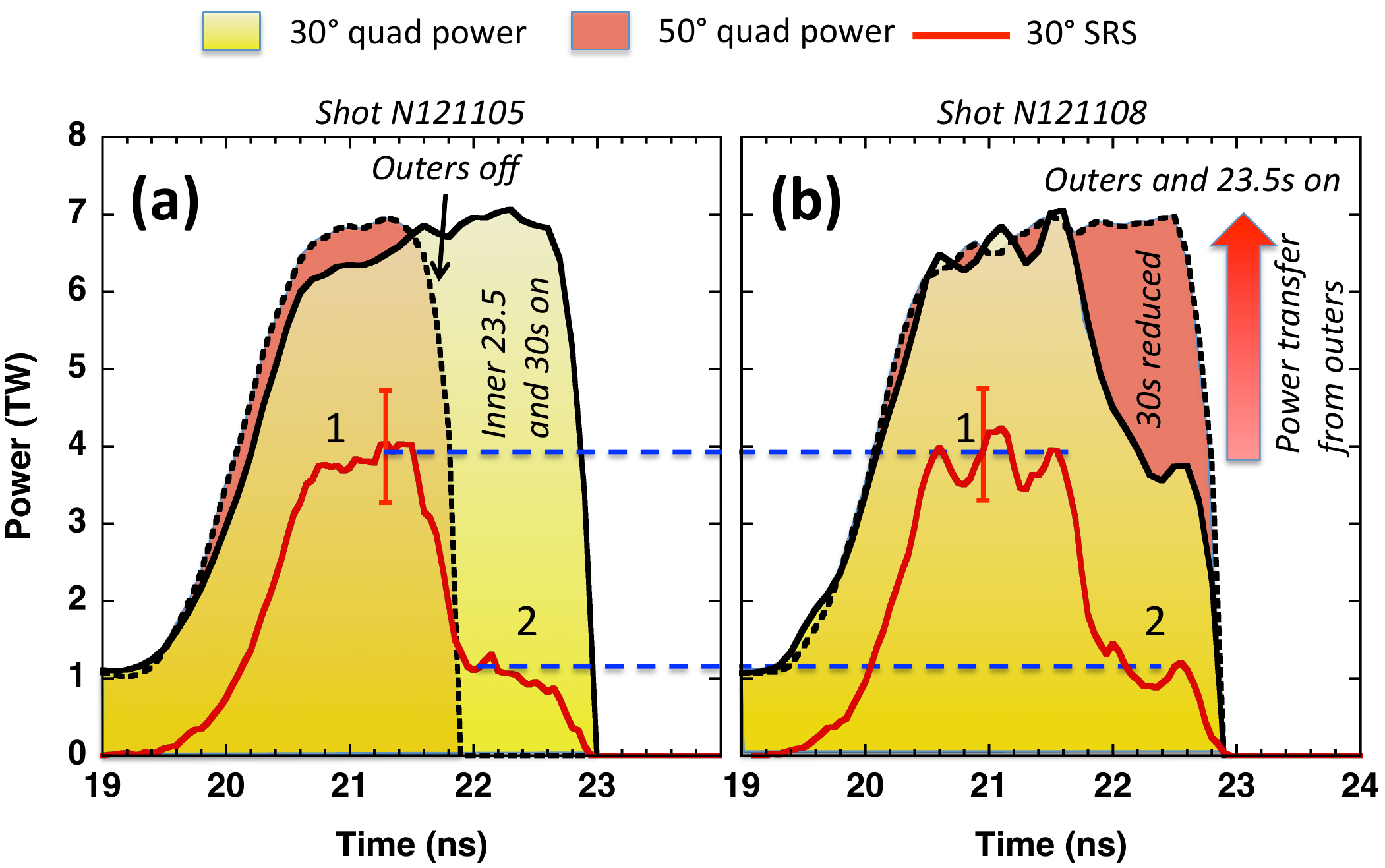} 
   \caption{(a) The outer cone beams turn off at 21.8 ns terminating the outer-inner power transfer and showing that the SRS is 1 TW when driven by the 30\degree quad only.  (b) Reducing the 30\degree power to about 55\% with power transfer active also achieves an SRS power of 1 TW.}
   \label{fig:fig_two}
\end{figure}

Quantification of the instantaneous power transfer from the outer to inner cones is made using the SRS and laser data shown in Fig.~\ref{fig:fig_two}.  The SRS reported is the total SRS for a single quad.  Figure \ref{fig:fig_two} (a) [NIF shot N121105] shows that in the epoch labeled "1", a measured SRS of 4 ($\pm$ 0.8) TW results for an incident laser power of 7 TW.  During this time, all quads are on and power transfer is active so the actual pump power is higher than 7 TW.  All outer quads turn off at about 21.8 ns causing power transfer from outers to inners to shut off and the SRS to drop to 1 ($\pm$ 0.2) TW in the epoch labeled "2."  The inner 23.5\degree quads are kept on so power transfer between inners is still active.  This measurement calibrates the 30\degree SRS backscatter to be 1 TW for a quad laser drive of
\begin{equation}
P_{2a}^{30} = P_0 \times [1 - \gamma_{30 \rightarrow 23.5}\times P_0],
\end{equation}
using Eq.~(1) and with $P_0 = 7$ TW.  In a second experiment shown in Fig.~\ref{fig:fig_two} (b) [shot N121108] the SRS once again decreases from 4 TW to 1 TW, but this time it is caused by a reduction in laser power on the 30\degree quads.  Using Eq.~(1) and that the power on both outer and the 23.5\degree cones is $P_0 = 7$ TW, the laser drive on the 30\degree quad in epoch "2" is
\begin{equation}
P_{2b}^{30} =  4{\rm TW}\times[1 + \gamma_{o \rightarrow i}\times P_0] [1 - \gamma_{30 \rightarrow 23.5}\times P_0 (1+\gamma_{o \rightarrow i}\times P_0)].
\label{eq:two}
\end{equation}
The key to using SRS as a laser power sensor for this data is the following: given that the measured SRS at the start of epoch 2 in the two Fig.~\ref{fig:fig_two} experiments is the same [1 ($\pm$ 0.2) TW], and the plasma properties at the start of epoch 2 in the two experiments are the same (see Fig.~3 discussion below), we conclude that the pump power must be the same in the 2 experiments.  Thus, $P_{2a}^{30}=P_{2b}^{30}$.  We measure a small transfer rate from the 30s to the 23.5s of $\gamma_{30 \rightarrow 23.5} = 0.01/{\rm TW} \pm 0.007$/TW for similar experiments changing the power on only the 23.5\degree or 30\degree cones.  This gives $ \gamma_{o \rightarrow i} = 0.125{\rm /TW} + 0.023{\rm /TW} - 0.013$/TW, and a total power increase to the 30\degree quads of 63\% $\pm$ 8.6\% when all other quad powers are 7 TW.  This measured range of transfer rates is consistent with {\it time-averaged} transfer rates inferred from the imploded core shape \cite{meezan,callahan}.
 
\begin{figure}[htbp]
   \centering
  \includegraphics[width=3.2in]{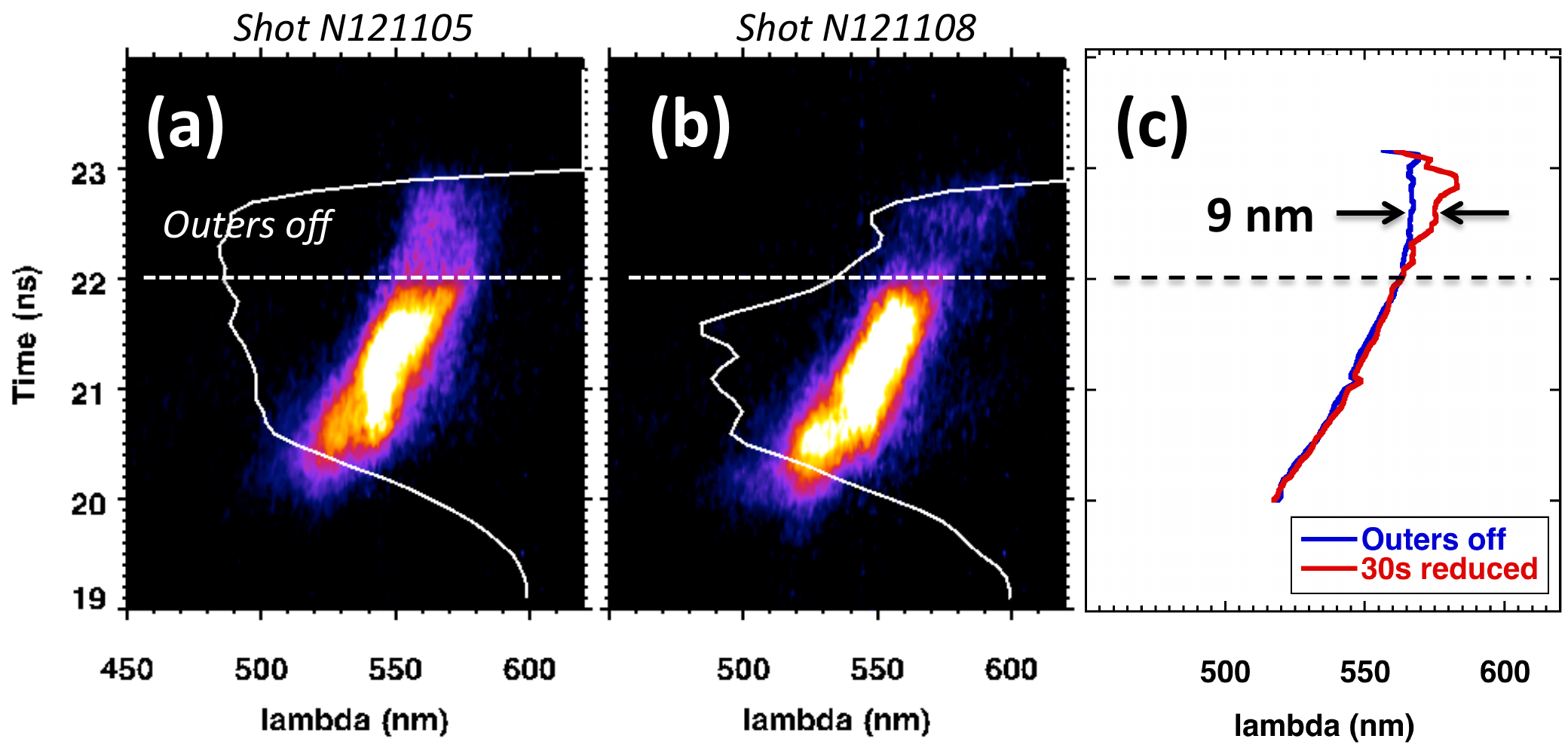} 
   \caption{(a) SRS streaked spectrum from the 30\degree FABS shows the drop in intensity when the outers turn off.  (b) SRS streaked spectrum for the case where the 30\degree cone drops to about one-half in intensity.  (c) average SRS wavelength (from the streaks) as a function of time showing that the wavelength for the case with outers off is about 9 nm shorter.  The thin white lines show the incident 30\degree quad power vs time.}
   \label{fig:fig_three}
\end{figure}

Figure \ref{fig:fig_three} (a) and (b) show the measured 30\degree FABS SRS spectra corresponding to the two experiments in Fig.~\ref{fig:fig_two}; the color scale has been adjusted to show the late time SRS better.  Figure \ref{fig:fig_three} (c) plots the wavelength centroid vs.\ time.  Immediately after the laser-power transition ($t=22$ ns) the SRS wavelengths in the two experiments are equal (within $\sim$ 1 nm around 560 nm) indicating that the plasma electron temperatures are equal to $\pm$ 0.1 keV.  Later in time the wavelengths separate reaching about 9 nm difference at 22.7 ns.  This wavelength separation corresponds to 0.3 to 0.4 keV using the Langmuir plasma wave dispersion relation \cite{trivel} and hydrodynamic simulations using Lasnex \cite{zimmer} are consistent with this estimate.  The incident laser and SRS absorption rate due to inverse bremsstrahlung is the same in both experiments up to and just after the transition time (where outers turn off or inners reduce power) and does not affect the determination of power transfer.  Measurements of SBS, energetic (10-100 keV) electrons, and hohlraum x-ray emission ($<$ 5 keV) are the same at $t=22$ ns (within measurement error) adding further confirmation that the hohlraum plasma is similar for the two Fig.~\ref{fig:fig_two} experiments.  Complimentary experiments using outer-beam SBS as a remote sensor of transfer from the outer quads \cite{aps2012}, show an outer-inner transfer rate consistent with the rate determined here using inner-beam SRS.  This indicates that the inner SRS power is determined dominantly by these transfer rates, and multi-beam effects or unmeasured side scatter which could lead to increased errors on the power transfer rates are not significant. Their possible secondary role is under investigation.

\begin{figure}[htbp]
   \centering
  \includegraphics[width=3.2in]{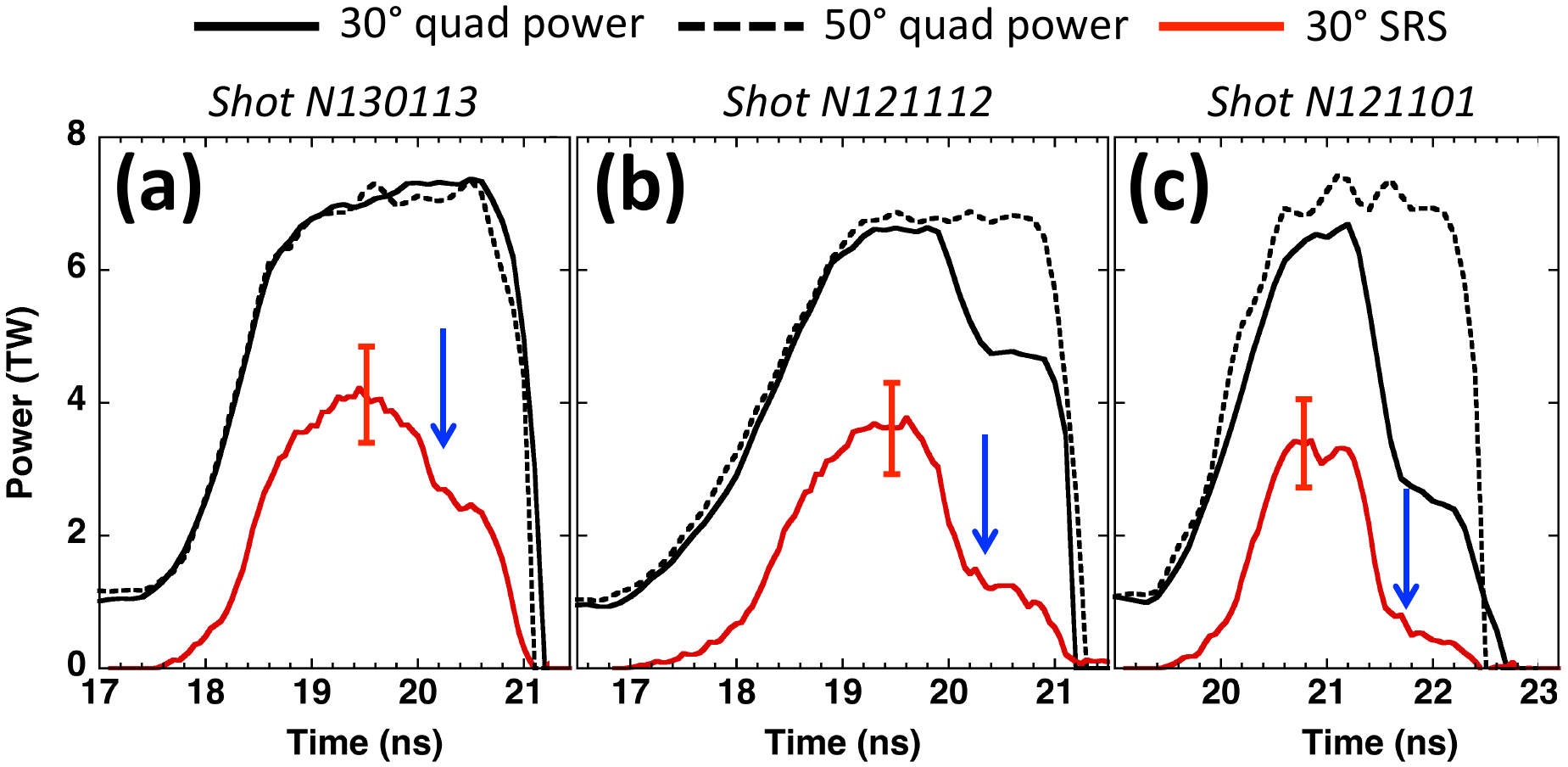} 
   \caption{Time-history of SRS for 3 cases: (a) both inners and outers remain on together, (b) the inner 30\degree quads drop to 75\% of the incident power with the outers on, and (c) the inner 30\degree quads drop to 40\% of the incident power with the outers on.  The blue arrows indicate times that the SRS shift is 560 nm $\pm$ 5 nm and SRS values used to construct Fig.~5.}
   \label{fig:fig_four}
\end{figure}

As an additional outcome of this transfer measurement we can now study SRS as a function of the post transfer laser power {\it at one point in time}.  Figure \ref{fig:fig_four} shows three additional SRS measurements made while "tuning" the drop in the 30\degree quad power to achieve an SRS power in epoch 2 of 1 TW as shown in Fig.~\ref{fig:fig_two} (b).  Figure \ref{fig:fig_four} (b) [shot N121112] shows the 30\degree power reduced to 75\% and (c) to 40\% [shot N121101].  Figure \ref{fig:fig_four} (a) [shot N130113] has both the inners and outers on throughout the pulse.  The timing of the three experiments is somewhat different due to the fact that the LPI measurements can be uniquely combined with hohlraum "shock-timing" measurements \cite{robey_one,robey_two}.  The SRS power dependence is determined at the times indicated by the arrows; this is just after the drop in the 30\degree quad power and when the SRS wavelength is at 560 nm $\pm$ 5 nm.  Since the plasma conditions are similar it's possible to isolate the SRS dependence on pump power at the time that the SRS wavelength is at $\sim$ 560 nm.  As noted previously, we use the measured cross-beam transfer rates for $\gamma_{o \rightarrow i}$ and $\gamma_{30 \rightarrow 23.5}$ obtained from the Fig.~\ref{fig:fig_two} data for similar plasma conditions to estimate the power-after-transfer for the Fig.~\ref{fig:fig_four} data.

\begin{figure}[htbp]
   \centering
  \includegraphics[width=3.2in]{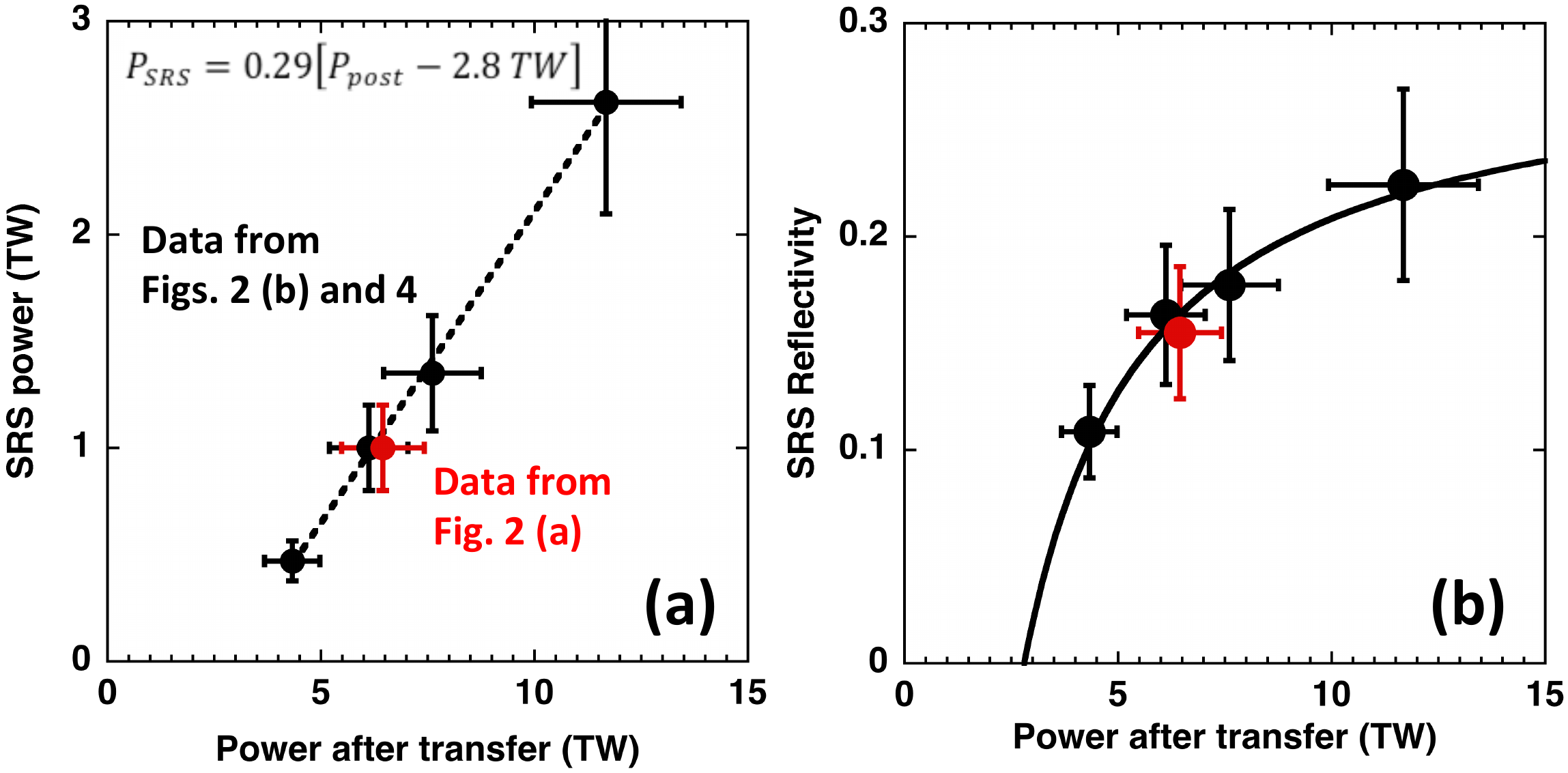} 
   \caption{(a) SRS backscatter power vs.\ laser drive after power transfer on a 30\degree quad.  The data lie on a line indicating a peak SRS reflectivity of 29\% of the incident power with a threshold of 2.8 TW.  (b) The same data shown as SRS reflectivity vs incident power.}
   \label{fig:fig_five}
\end{figure}

Figure \ref{fig:fig_five} (a,b) plots the SRS (power, reflectivity) vs. the laser power after transfer for the three experiments in Fig.~\ref{fig:fig_four} and the two in Fig.~\ref{fig:fig_two}.  The error bars account for SRS measurement and power-transfer uncertainty.  The red data point is from Fig.~\ref{fig:fig_two} (a) where the outer beams were turned off, setting the outer power transfer to zero. The data points are fit well by the expression:
\begin{eqnarray}
\Psrs  = \Rmax\times[\Ppost - \Pthr]
\label{eq:three}.
\end{eqnarray}
$\Rmax=0.29$ is the maximum reflectivity \cite{reflectivity}, $\Pthr=2.8$ TW is the threshold power and $\Ppost$ is the post-transfer power.  This has the same functional form as the large gain limit of the 1D Tang solution to the strongly-damped coupled-mode equations \cite{tang}.  

In summary, we have measured for the first time the instantaneous outer-to-inner cone cross-beam energy transfer using SRS as a calibrated remote laser-power sensor.  We find that the inner 30\degree quads increase by 63\%.  A continuing aspect of this study is to map out the cross-beam transfer at different times in the pulse.

We acknowledge the help of Dr. R. L. Berger.  This work was performed under the auspices of the U.S. Department of Energy by Lawrence Livermore National Laboratory under Contract DE-AC52-07NA27344.

\end{document}